\input harvmac
\overfullrule=0pt
\font\authorfont=cmcsc10 \ifx\answ\bigans\else scaled\magstep1\fi
{\divide\baselineskip by 4
\multiply\baselineskip by 3

\Title{hep-th/9808157}{\vbox{\centerline{Yang-Mills 
Instantons in the Large-$N$ Limit} \bigskip  
\centerline{and the AdS/CFT Correspondence}}}
\vskip2pt
\centerline{\authorfont Nicholas Dorey$^{1,2}$, Valentin V. Khoze$^3$,
 Michael P. Mattis$^4$, {\rm and} Stefan Vandoren$^1$}
\bigskip
\centerline{\sl $^1$Physics Department, University  of Wales Swansea}
\centerline{\sl Swansea SA2$\,$8PP UK }
\bigskip
\centerline{\sl $^2$Physics Department, University of Washington}
\centerline{\sl Seattle, WA 98195 USA}
\bigskip
\centerline{\sl $^3$Department of Physics, Centre for Particle Theory, 
University of Durham}
\centerline{\sl Durham DH1$\,$3LE UK }
\bigskip
\centerline{\sl $^4$Theoretical Division T-8, Los Alamos National Laboratory}
\centerline{\sl Los Alamos, NM 87545 USA}
\vskip .3in
\noindent
We examine a certain 16-fermion correlator in ${\cal N}=4$ supersymmetric
$SU(N)$ gauge theory in 4 dimensions. Generalizing recent $SU(2)$ results of
Bianchi, Green, Kovacs and Rossi, we calculate the exact $N$-dependence of
the effective 16-fermion vertex at the 1-instanton level, 
and find precise agreement in the
large-$N$ limit with the prediction of the type IIB superstring on
AdS$_5\times S^5.$ This suggests that the string theory prediction
for the 1-instanton amplitude considered here is not corrected by
higher-order terms in the $\alpha'$ expansion.
\vfil\break
}
\def\bigZ{Z\!\!\!Z}
\def\bigR{{\rm I}\!{\rm R}}
\def\sigmabar{\bar\sigma}
\def\mubar{\bar\mu}
\def\dalpha{{\dot\alpha}}
\def\dgamma{{\dot\gamma}}
\def\etabar{\bar\eta}
\def\Sinst{S_{\rm inst}}
\def\Squad{S_{\rm quad}}
\def\N{{\cal N}}
\def\M{{\cal M}}
\def\gst{g_{st}}
\lref\PP{ J. Polchinski and P. Pouliot, {\it Phys. Rev.} {\bf D56} (1997) 
6601, hep-th/9704029.}
\lref\BFSS{ T. Banks, W. Fischler, N. Seiberg and L. Susskind, 
{\it Phys. Lett.} {\bf B408} (1997) 111, hep-th/9705190.}
\lref\SS{S. Paban, S. Sethi and M. Stern, {\it  ``Summing up Instantons 
in Three-Dimensional Yang-Mills Theories''}, {hep-th/9808119}.}
\lref\AG{L. Alvarez-Gaum$\acute{\rm e}$,
 {\it Commun. Math. Phys.} {\bf 90} (1983) 161.}
\lref\ADHM{M.  Atiyah, V.  Drinfeld, N.  Hitchin and
Yu.~Manin, Phys. Lett. A65 (1978) 185. }
\lref\tHooft{G. 't Hooft, Phys. Rev. D14 (1976) 3432; ibid.
(E) D18 (1978) 2199.}
\lref\CGTone{ E. Corrigan, D. Fairlie, P. Goddard and S. Templeton,
    Nucl. Phys. B140 (1978) 31; 
E. Corrigan, P. Goddard and S. Templeton,
Nucl. Phys. B151 (1979) 93.}
\lref\Bernard{C. Bernard, Phys. Rev. D19 (1979) 3013.    }
\lref\AFone{L. Andrianopoli and S. Ferrara, \it
$K-K$ Excitations on  $ADS(5) \times S^5$ as $N=4$ ``Primary'' Superfields,
\it Phys. Lett. \bf B430 \rm (1998) 248,
 hep-th/9803171.}
\lref\FFone{S. Ferrara and C. Fronsdal, 
\it Conformal Maxwell Theory as a Singleton Field Theory
on $ADS(5)$, IIB Three-Branes, and Duality,
\it Class. Quant. Grav. \bf15 \rm (1998) 2153,
hep-th/9712239.}
\lref\FMMR{D.Z. Freedman, S.D. Mathur, A. Matusis and L. Rastelli,
\it Correlation Functions in the $CFT(D)/ADS(D+1)$ Correspondence\rm,
hep-th/9804058; and hep-th/9808006. }
\lref\measone{N. Dorey, V. Khoze and M. Mattis, \it Supersymmetry and
the multi-instanton measure\rm,
\it Nucl. Phys. \bf B513 \rm (1998) 681,
 hep-th/9708036.}
\lref\kms{V. Khoze, M. Mattis and M. Slater, \it The Instanton
Hunter's Guide to $SU(N)$ SUSY Gauge Theories\rm, \it Nucl. Phys. \bf
B \rm (to appear),
hep-th/9804009.}
\lref\Cordes{S. Cordes, Nucl. Phys. B273 (1986) 629.  }
\lref\WB{ J. Wess and J. Bagger, {\it Supersymmetry and Supergravity}, 
Princeton University Press, 1992.
  }
\lref\meastwo{N. Dorey, T. Hollowood,
V. Khoze and M. Mattis, \it Supersymmetry and
the multi-instanton measure II: From N=4 to N=0\rm, 
\it Nucl. Phys. \bf B519 \rm (1998) 470,
hep-th/9709072.}
\lref\MV{W. Muck and K. S. Viswanathan, hep-th/9804035 and 9805145; 
G. Chalmers, H. Nastase, K. Schalm and R. Siebelink, hep-th/9805105.}
\lref\dkmone{N. Dorey, V.V. Khoze and M.P. Mattis, \it Multi-instanton
calculus in $N=2$ supersymmetric gauge theory\rm, hep-th/9603136,
Phys.~Rev.~D54 (1996) 2921.}
\lref\dkmfour{N. Dorey, V.V. Khoze and M.P. Mattis, \it Multi-instanton
calculus in $N=2$ supersymmetric gauge theory.
II. Coupling to matter\rm, hep-th/9607202, Phys.~Rev.~D54 (1996) 7832.}
\lref\Amati{D. Amati, K. Konishi, Y. Meurice, G. Rossi and G. Veneziano,
Phys. Rep. 162 (1988) 169.}
\lref\FP{D. Finnell and P. Pouliot,
{\it Instanton calculations versus exact results in 4 dimensional 
SUSY gauge theories},
Nucl. Phys. B453 (95) 225, hep-th/9503115.}
\lref\GGK{M. B. Green, M. Gutperle and H. Kwon, \it Sixteen-fermion and
related terms in M theory on $T^2,$ 
\it Phys. Lett. \bf B421 \rm (1998) 149,
\rm hep-th/9710151.}
\lref\Maldacena{J. Maldacena, {\it ``The Large $N$ Limit of Superconformal Field 
Theories and Supergravity''}, {\rm hep-th/9711200}.  }
\lref\KT{S. Gubser, I. Klebanov and A. Peet, {\it Phys. Rev.} 
{\bf D54} (1996) 3915, hep-th/9602135; 
 I. Klebanov and A. A. Tseytlin, {\it Nucl. Phys.} {\bf B475} (1996) 179,
hep-th/9604166;
 S. Gubser, I. Klebanov, {\it Phys. Lett.} {\bf B413} (1997) 41
hep-th/9708005. }
\lref\BGKR{M. Bianchi, M. B. Green, S. Kovacs and 
G. Rossi,  {\it ``Instantons in Supersymmetric Yang-Mills theory and 
D-instantons in IIB Superstring Theory''}, {\rm hep-th/9807033}. }
\lref\GKP{S. Gubser, I. Klebanov and A. M. Polyakov, {\it ``Gauge Theory 
Correlators from Non-Critical String-Theory''},
\it Phys. Lett. \bf B428 \rm (1998) 105,
 {\rm hep-th/9802109}. }
\lref\HS{M. Henningson and K. Sfetsos, {\it ``Spinors and the AdS/CFT 
correspondence''}, {\rm hep-th/9803251}. }
 \lref\Witten{E. Witten, {\it ``Anti-de-Sitter Space and Holography''}, 
{\rm hep-th/9802150}. }
\lref\dkmthreed{N. Dorey, V. V. Khoze and M. P. Mattis, 
\it  Multi-Instantons, Three-Dimensional Gauge Theory, and
the Gauss-Bonnet-Chern Theorem\rm, 
{\it Nucl. Phys.} 
{\bf B502} (1997) 94, hep-th/9704197. }
\lref\dkmdef{N. Dorey, V. V. Khoze and M. P. Mattis, 
\it On Mass-Deformed
N=4 Supersymmetric Yang-Mills Theory\rm,
{\it Phys. Lett.} 
{\bf B396} (1997) 141, hep-th/9612231. }
\lref\BG{T. Banks and M. B. Green, 
{\it ``Non-Perturbative Effects in $AdS_{5}\times S^{5}$ 
String Theory and $d=4$ SUSY Yang-Mills''}, {\rm hep-th/9804170}. }

In interesting recent work by Maldacena \Maldacena\ (see also 
\KT\ for relevant previous work by other authors), 
the large-$N$ limit of ${\cal N}=4$ supersymmetric 
Yang-Mills theory (SYM) has been related to the low-energy behavior of 
type IIB superstrings on AdS$_{5}\times S^{5}$. 
In the conjectured correspondence, the 
gauge coupling $g$ and vacuum angle $\theta$
of the four-dimensional theory are given by
\eqn\corresp{g\ =\ \sqrt{4\pi\gst}\ =\ \sqrt{4\pi e^\phi}\ ,\qquad
\theta\ =\ 2\pi c^{(0)}}
Here $g_{st}$ is the string coupling while
$c^{(0)}$ is the expectation value of the Ramond-Ramond scalar of IIB 
string theory. Also $N$ appears explicitly, through the relation
\eqn\alphadef{{L^2\over\alpha'} \ =\ \sqrt{g^2\,N}}
where $L$
 is the radius of both the AdS$_5$ and $S^5$ factors of the background.
One striking consequence of these identifications is 
that the action of a Yang-Mills instanton in the gauge theory is equated 
to that of a D-instanton in the string theory. The relation between 
these two seemingly different types of instantons has been investigated 
further by Bianchi, Green, Kovacs and Rossi (BGKR) \BGKR. 
These authors compared the leading semiclassical contribution 
of a single Yang-Mills instanton in the ${\cal N}=4$ theory with gauge group 
$SU(2)$ with that of a D-instanton in the 
IIB theory on AdS$_{5}\times S^{5}$. Specifically, the gauge theory is 
at its conformal point where all the Higgs VEVs vanish and the instanton is 
an exact solution of the field equations.  BGKR 
found an interesting correspondence between the moduli-space and 
zero modes of the two classical configurations and between the resulting 
contributions to various correlation functions in their 
respective theories. A particularly 
attractive aspect of the correspondence is that the 
scale size of the Yang-Mills instanton is mapped onto the radial position of 
the D-instanton in AdS$_{5}$. This is consistent with the interpretation 
of Maldacena's conjecture in which the four-dimensional gauge theory 
lives on the boundary of AdS$_{5}$ \refs{\GKP,\Witten}.  

An obvious puzzle about the results of Ref.~\BGKR\ is that 
such an agreement between weakly-coupled 
gauge theory and the low-energy effective field theory 
of the IIB string is found for gauge group $SU(N)$ with $N=2$.  
In contrast, Maldacena's conjecture only predicts such an agreement 
in the large-$N$ limit.\foot{More accurately, as explained below, the 
low-energy IIB prediction should only hold when $g^{2}N$ is large. 
On the other hand, to 
justify using semiclassical methods we must also have $g^{2}$ small. 
These conditions together certainly require $N$ to be large.}
In this letter we generalize to arbitrary $N$ 
the $SU(2)$ calculation of the 1-instanton contribution to 
a sixteen fermion correlator considered in \BGKR. In the $SU(N)$ SYM theory 
a single instanton has a total of $8N$ adjoint fermion 
zero modes and, for $N=2$, the resulting Grassmann integrations are 
saturated by the sixteen fermion insertions of this correlator. 
For $N>2$ there are additional fermion zero modes which must be 
lifted in order to obtain a non-zero result. Thus  
the main technical challenge in generalizing the calculation of BGKR 
is to account correctly for this lifting. As in several other cases in  
three \dkmthreed\ and four \dkmdef\ dimensions, this can be 
accomplished by determining the Grassmann quadrilinear term in the 
instanton action.   Our final result for the instanton contribution 
has a complicated algebraic dependence on $N$. However, 
in the large-$N$ limit, the dependence on both $g^{2}$ 
and $N$ is exactly that extracted from the superstring 
by BGKR, and earlier
 by Banks and Green \BG, on the basis of 
Maldacena's conjecture. 
 
It is important to emphasize that, although 
we are working in the large-$N$ limit, the agreement we have found 
is still somewhat mysterious. Maldacena's conjecture 
relates the ratio of the string length-scale, $\alpha'$, 
and the radius of curvature of the 
spacetime, $L$, to the gauge theory parameters via Eq.~\alphadef.
This means that the $\alpha'$ expansion of the 
IIB theory, on which the prediction of Refs.~\BGKR\ and \BG\ relies, 
is only valid in the regime of large $g^{2}N$. 
In particular, it is not obvious that stringy corrections 
to the low-energy IIB effective action 
(i.e., higher orders in the $\alpha'$ expansion) 
can be neglected when comparing to our semiclassical calculation.   
A similar situation arises for the calculation of  
an eight-fermion correlator in the three-dimensional theory with sixteen 
supercharges \refs{\PP,\dkmthreed}, 
where {\it weak}-coupling multi-instanton calculations agree 
exactly with the predictions 
based on the M(atrix) model of M-theory \refs{\PP,\BFSS}
which, strictly speaking, only apply 
in a {\it strong}-coupling limit. As emphasized by Banks and Green, in
general such an agreement would 
only be expected if the relevant correlator is constrained by a 
supersymmetric nonrenormalization theorem. 
Very recently 
 \SS, exactly such a non-renormalization theorem has been proved 
for eight-fermion terms in the effective action of the 
three-dimensional theory. Our present results suggest that a similar 
nonrenormalization theorem should be at work in the 
four-dimensional context. Specifically, it appears that 
the prediction for the 16 fermion correlator extracted from the 
low-energy IIB action in \refs{\BG,\BGKR} is not modified by 
higher-order stringy corrections. 
On the other hand, the exact $N$-dependence of our one-instanton result 
given below
includes an infinite series of $1/N$ corrections which should correspond to
quantum corrections on the IIB side. 

\def\calG{{\cal G}}
We first briefly review the type IIB superstring prediction, closely following
\BGKR. At leading order beyond the Einstein-Hilbert term
in the derivative expansion, the IIB effective action
is expected to contain a totally antisymmetric 16-dilatino effective vertex
of the form \refs{\GGK,\BG}
\eqn\effvert{(\alpha')^{-1}\int d^{10}X\,\sqrt{\det g}\,e^{-\phi/2}\,
f_{16}(\tau,\bar\tau)\,\Lambda^{16}\ + \ \hbox{H.c.}}
Here $\Lambda$ is a complex chiral $SO(9,1)$ spinor, and $f_{16}$ is a certain
weight $(12,-12)$ modular form under $SL(2,\bigZ)$. 
In particular $f_{16}$ has the following weak-coupling expansion:
\eqn\fexpand{f_{16}\ =\ a_0\,e^{-3\phi/2}+a_1\,e^{\phi/2}
+\sum_{k=1}^\infty\calG_k\,e^{\phi/2}\ ,}
where
\eqn\Gdef{\calG_k\ =\ \Big(\sum_{n|k}{1\over n^2}\Big)\,\big(ke^{-\phi})^{25/2}\,
\exp\big(-2\pi k( e^{-\phi}+ic^{(0)})\,\big)}
neglecting perturbative corrections; the summation in \Gdef\ runs over the positive
integral divisors of $k$. Notice that with the conjectured correspondence \corresp\
to the couplings of 4D SYM theory, the expansion \fexpand\ has the
structure of a semiclassical expansion: the first two terms correspond to
the tree and one-loop pieces, respectively ($a_0$ and $a_1$
 are numerical constants),
while the sum on $k$ is interpretable as a sum on Yang-Mills instanton number.
In the IIB theory, these terms, which are non-perturbative in the string 
coupling come from D-instantons.

{}From this effective vertex one can construct Green's functions for 16 
fermions $\Lambda(x_i),$ $1\le i\le 16,$ which live on the boundary
of AdS$_5$:
\eqn\grnfcn{\langle\,\Lambda(x_1)\cdots\Lambda(x_{16})\rangle\ \sim\
(\alpha')^{-1}\,e^{-\phi/2}\,f_{16}\,t_{16}\int{d^4x_0\,d{\rho}\over{\rho}^5}\,
\prod_{i=1}^{16}\,K_{7/2}^F(x_0,{\rho};x_i,0)}
suppressing spinor indices.
Here $K_{7/2}^F$ is the bulk-to-boundary propagator for a spin-$1/2$ Dirac
fermion
of mass $m=-3/2L$ and scaling dimension $\Delta=7/2\,$
\refs{\HS,\GKP,\Witten,\FMMR}:
\eqn\KFdef{K_{7/2}^F(x_0,{\rho};x,0)\ =\ 
K_{4}(x_0,{\rho};x,0)\,\big(\rho^{1/2}\gamma_5+\rho^{-1/2}(x_0-x)_n
\gamma^n\big)
}
with
\eqn\Kfourdef{K_{4}(x_0,{\rho};x,0)\ =\ {{\rho}^4\over\big({\rho}^2+
(x-x_0)^2\big)^4}}
In these expressions the $x_i$ are 4-dimensional space-time coordinates for the
boundary of AdS$_5$ while $\rho$ is the fifth, radial, coordinate; we suppress
the coordinates on $S^5$ as the propagator does not depend on them (save through
an overall multiplicative factor which we drop). The quantity $t_{16}$
in Eq.~\grnfcn\ is (in the notation of BGKR) a 16-index antisymmetric
invariant tensor which enforces Fermi statistics and ensures, \it inter
alia\rm, that precisely 8 factors of $\rho^{1/2}\gamma_5$ and
8 factors of $\rho^{-1/2}\gamma^n$ are picked out in the product over
$K^F_{7/2}.$

According to Maldacena's conjecture, the  correlator
\grnfcn\ in the IIB theory should correspond to a certain 16-fermion 
correlator in 4D large-$N$ SYM. 
The correspondence of operators in the two pictures
was established in Refs.~\refs{\GKP,\Witten,\AFone,\FFone}.
 For present purposes,
 the fermion operator in the 4D SYM picture
with the right transformation properties is the gauge invariant
composite operator
\eqn\compop{\Lambda_\alpha^A\ =\ \sigma^{mn}{}_\alpha^{\ \beta}\,\Tr_N(
v_{mn}\,\lambda_\beta^A)}
which is a spin-$1/2$ fermionic Noether current associated with a particular
superconformal transformation. Here $v_{mn}$ is the  $SU(N)$
gauge field strength while the $\lambda_\beta^A$ are the 
Weyl gauginos, with the index $A=1,2,3,4$ labeling the four 
supersymmetries. The numerical tensor $\sigma^{mn}$ projects out the
self-dual component of the field strength,\foot{We use Wess and
Bagger conventions throughout \WB.} so that only  instantons
rather than  anti-instantons can contribute.

In Ref.~\BGKR, BGKR explicitly compared the form of the first term in the 
D-instanton expansion of the 16 dilatino correlator  \grnfcn\ with 
the 1-instanton contribution to a correlator in ${\cal N}=4$ $SU(2)$ 
Yang-Mills theory with 16 insertions of the operator \compop.        
These authors noted that the two correlators agree up to an overall 
normalization. In particular the integration measure and integrand appearing 
in Eq.~\grnfcn\ exactly match their counterparts in 
the gauge theory calculation. As reviewed below, 
technically this identification is possible because
the expression \Kfourdef\  is proportional to the
1-instanton action density:
\eqn\works{\tr_N(v_{mn})^2\,{\Big|}_{\rm 1\hbox{-}inst}\ 
=\ {96\over g^2}\,K_4\ .}
As mentioned earlier, for gauge group $SU(2)$ the single $\N=4$ superinstanton
 contains precisely 16 adjoint
fermion zero modes: a supersymmetric plus a superconformal
zero mode, each of which is a Weyl 2-spinor, times four supersymmetries.
A nonvanishing 16-fermion correlator is therefore obtained by saturating
each of the fermion insertions with a distinct such zero mode. 
In what follows we generalize the calculation of BGKR to $SU(N)$. Since for
any value of $N$
the relation \works\ still works,
 the functional similarity between
the 16-fermion correlators in the two pictures noted by BGKR
continues to hold.
 However, by correctly accounting for the $8N-16$ lifted
fermion modes,
we will also extract the overall multiplicative
constant $C_N$ which determines the strength of the effective
16-fermion SYM vertex.
 As shown
below, in the large-$N$ limit, $C_N$ scales like $\sqrt{N}.$ We
identify this behavior with the factor of $1/\alpha'$ in front of the
IIB effective vertex \effvert, using the dictionary
\alphadef. The extraction of this overall constant as a function of $N$
is our chief result. 

In order to calculate the required 16-fermion  correlator in 4D SYM theory,
%
%
one needs to understand
(i) the collective coordinate integration measure, (ii) the instanton action
$\Sinst,$ and (iii) the form of the fermionic insertions \compop. Let us
discuss each of these, in turn:

\bf (i) The measure\rm. For general topological number $k$, the collective
coordinate integration measure in the $\N=1,$ $\N=2$ and $\N=4$ 
supersymmetric cases was derived in
Refs.~\refs{\measone\meastwo-\kms}.\foot{The one specific case of the
$\N=4$ $k$-instanton measure for general $SU(N)$ is not explicitly
given in these references, but may be written down by inspection,
by generalizing the $SU(2)$ expression (7) of Ref.~\meastwo\ to $SU(N)$ by the
methods of Ref.~\kms. When $k=1$, this measure reduces to
Eq.~(13) below.} The form of these measures is uniquely fixed
by supersymmetry, the index theorem, renormalization group decoupling,
and cluster decomposition. The collective coordinates used in these
expressions are 
those of the ADHM multi-instanton \refs{\ADHM,\CGTone},
 suitably supersymmetrized as explained
in Refs.~\refs{\dkmone,\dkmfour,\kms}. In particular the bosonic and
adjoint fermionic collective coordinates are encoded in ADHM matrices
$a$ and $\M^A,$ respectively, where the index $A$ runs over the independent
supersymmetries. For gauge group $SU(N)$ and topological number $k$,
$a$ is an $(N+2k)\times2k$ complex-valued matrix, while $\M^A$ is an
$(N+2k)\times k$ matrix of complex Grassmann numbers (see Ref.~\kms\ for
a review). The elements of $a$ and $\M^A$ are subject to polynomial constraints
and gauge-like invariances which reduce the number of independent collective
coordinates to the number required by the index theorem.
In the 1-instanton sector these matrices have the simple canonical 
form \kms:
\eqn\matdef{a\ =\ \pmatrix{0\quad 0\cr\vdots\quad \vdots\cr0\quad 0\cr\rho
\quad 
0\cr0\quad \rho\cr
-x_0^m\sigma_m\cr}\ ,\quad\M^A\ =\ \pmatrix{\mu^A_1\cr\vdots\cr\mu^A_{N-2}\cr
4i\rho\etabar^{A1}\cr 4i\rho\etabar^{A2}\cr4\xi_1^A\cr4\xi_2^A}}
We have chosen familiar notation whereby $\rho\in\bigR$ and $x_0^m\in
\bigR^4$ denote the size and position of the instanton, and $\xi_\alpha^A$
and $\etabar^{A\dot\alpha}$ are the supersymmetric and superconformal
fermion zero modes, respectively.
 Equation \matdef\ assumes the
canonical `North pole' embedding of the instanton within $SU(N)$;
more generally there is a manifold of equivalent instantons obtained
by acting on \matdef\ by group generators $\Omega$ in the coset space
\eqn\cosetdef{\Omega\ \in\ {U(N)\over U(N-2)\times U(1)}}
The complex Grassmann coordinates $\mu_i^A$ in
Eq.~\matdef\ (which do not
carry a Weyl spinor index) may be thought of as the superpartners of the coset
embedding parameters \cosetdef.
 Together, $\xi_\alpha^A,$ $\etabar_\dalpha^A,$
$\mu_i^A$ and $\mubar_i^A$ constitute $8N$ fermionic 
collective coordinates, as needed.

In terms of these 1-instanton variables,
the correctly normalized
$\N=4$ collective coordinate integration measure may be easily deduced
from the early literature 
\refs{\tHooft\Bernard-\Cordes}:\foot{The normalization factors of $\xi^A$
and $\etabar^A$,
$g^2/16\pi^2\mu_0$ and $g^2/32\pi^2\rho^2\mu_0$ respectively, used
in this measure, agree with 
Refs.~\refs{\FP,\dkmone} but disagree, by factors of two, with
Ref.~\Amati\ and much of the subsequent literature; 
see Ref.~\FP\ for a discussion. In the ADHM language, the relative values
of the $\xi^A,$ $\etabar^A$ and $\mu^A$ normalizations used in this
measure follow from Eq.~\matdef\ together with the fact that the
inner product of ${\cal M}^A$ matrices is proportional to
$\Tr\bar\M^A({\cal P}_\infty+1)\M^A$ where the $(N+2k)\times(N+2k)$ diagonal
matrix ${\cal P}_\infty+1$ has 2's in the first $N$ diagonal entries
and 1's in the remaining $2k$ diagonal entries \refs{\CGTone,\dkmone}.
}
\eqn\measdef{\eqalign{&{2^{4N+2}\pi^{4N-2}\over(N-1)!(N-2)!}\,{1\over g^{4N}}
\int d^4x_0\int{d\rho\over\rho^5}(\mu_0\rho)^{4N}
\cr&\times\ 
\int\prod_{A=1}^4 d^2\xi^A\,\big({g^2\over16\pi^2\mu_0}\big)^4
\int\prod_{A=1}^4 d^2\etabar^A\,\big({g^2\over32\pi^2\rho^2\mu_0}\big)^4
\int\prod_{A=1}^4\prod_{i=1}^{N-2}\,d\mu_i^A d\mubar_i^A\,
\big({g^2\over2\pi^2\mu_0}\big)^{4(N-2)}}}
As usual in instanton calculations \tHooft, $g=g(\mu_0)$ is evaluated in
the Pauli-Villars scheme; since the $\N=4$ model is a finite theory,
the subtraction scale $\mu_0$ should, and by inspection does, cancel out of
Eq.~\measdef.
The overall constant in front (see Eq.~(33) of \Bernard) comes from the 
volume of the coset space \cosetdef;
this factor presupposes that the measure will be used to integrate
only gauge singlets, as we shall be doing (since all adjoint Higgs VEVs,
which pick out special directions in the color space,
will be set to zero in the present paper).

\bf (ii) The instanton action\rm. For all instanton numbers $k$, the
$\N=4$ instanton action was derived in Ref.~\dkmdef, for the gauge group
$SU(2)$. Using the methods of Ref.~\kms, that expression immediately
generalizes to $SU(N)$ for arbitrary $N$. In particular, in
the conformal case where
all adjoint Higgs VEVs are set to zero, the instanton action has the form
\eqn\instact{\Sinst\ =\ {8\pi^2k\over g^2}+\Squad}
Here $\Squad$ is a particular fermion quadrilinear term, with one
fermion collective coordinate chosen from each of
the four gaugino sectors $A=1,2,3,4.$
In the 1-instanton sector this term collapses to\foot{See Eqs.~(14), (4)
and (6) of \dkmdef, and Eqs.~(3.20), (8.6) and (8.7) of \kms.}
\eqn\squadef{\Squad\ =\ {\pi^2\over2\rho^2g^2}\,\epsilon^{}_{ABCD}\,
\Lambda^N_f(\M^A,\M^B)\Lambda^N_f(\M^C,\M^D)}
where
\eqn\Lambdafdef{\Lambda^N_f(\M^A,\M^B)\ =\ {1\over2\sqrt{2}}\,
\sum_{i=1}^{N-2}\big(\mubar_i^A\mu_i^B-\mubar_i^B\mu_i^A\big)}
As we explained in Ref.~\dkmdef, for all $k$, in the absence of VEVs,
$\Squad$ is a supersymmetric invariant quantity that lifts all the
fermion zero modes except for the 16 supersymmetric and superconformal
modes. For $k=1$ this latter property is obvious
from Eq.~\Lambdafdef, which explicitly depends only on the collective
coordinates $\mu_i^A$
and $\mubar_i^A$ and not on  $\xi_\alpha^A$ or $\etabar_\dalpha^A$.

\bf (iii) The fermion insertions\rm. 
As stated earlier, the 16 explicit fermion insertions \compop\
are needed to saturate the 16 global supersymmetric and superconformal zero
modes, which are not otherwise lifted by the action 
\squadef. Accordingly we substitute for the gaugino $\lambda_\beta^A$ in
Eq.~\compop:
\eqn\gnomode{\lambda_\beta^A(x)\ =\ 
-\big(\xi^{A\gamma}-\etabar_\dgamma^A\sigmabar_m^{\dgamma\gamma}
\cdot(x^m-x_0^m)\big)
\sigma^{kl}_{\gamma\beta}\,v_{kl}(x-x_0)+\cdots}
as follows from Eqs.~(4.3a) and (A.5) of \dkmone;
 the dots  stand for admixtures of the
remaining fermion modes which we can neglect (since these are saturated
by $\Sinst$). The field strengths in
Eqs.~\compop\ and \gnomode\ are to be evaluated on the instanton.
With Eq.~\works\ together with the identity
\eqn\projid{\tr_N\,v_{mn}v_{kl}\ =\ {1\over3}\,{\cal P}_{mn,kl}^{\rm SD}\,
\tr_N(v_{pq})^2\ ,}
where ${\cal P}_{mn,kl}^{\rm SD}$ is the projector onto self-dual
antisymmetric tensors, Eqs.~\compop\ and \gnomode\ reduce to
\eqn\insertid{\eqalign{\Lambda^{A\gamma}(x)\ &=\ -
\big(\xi^{A\gamma}-\etabar_\dgamma^A\sigmabar_m^{\dgamma\gamma}
\cdot(x^m-x_0^m)\big)
\tr_N\big(v_{pq}(x-x_0)\big)^2
\cr& =\ 
-{96\over g^2}\,\big(\xi^{A\gamma}-\etabar_\dgamma^A\sigmabar_m^{\dgamma\gamma}
\cdot(x^m-x_0^m)\big)
K_4(x_0,\rho;x,0)}}

Putting this all together, we now insert 16 copies of the composite
fermion \insertid, at 16 spacetime points $x_i,$ times $\exp(-\Sinst)$,
into the 1-instanton collective coordinate measure \measdef. Apart
from the integrations over the lifted fermion modes $\{\mu_i^A,
\mubar_i^A\}$ and the \hbox{$N$-dependent} constant in front of the measure,
the resulting expression is identical (up to some corrected factors
of 2) to the $SU(2)$ expression examined by BGKR. Our task is precisely
to evaluate
the $N$-dependent
effect of the lifted modes. 
On a formal level, 
for general $N$, and general topological number $k$, the integral of
$\exp(-\Squad)$ over the lifted fermion modes is given by the
Gauss-Bonnet-Chern theorem, and  equals the Euler character
of a certain quotient space formed from the charge-$k$ $SU(N)$ ADHM
moduli space \refs{\dkmthreed,\AG}. However, since much less is
known about such spaces than about the multi-monopole spaces which
govern analogous instanton calculations in 3D \dkmthreed, here we
will adopt an alternative, direct calculational 
approach. Specializing to $k=1,$
we define $I_N$ to be the (unnormalized) 
contribution of the lifted modes to the correlator:
\eqn\INdef{I_N\ =\ \int\prod_{A=1}^4\prod_{i=1}^{N-2}d\mu_i^Ad\mubar_i^A
\,e^{-\Squad}}
By explicit computation we find
\eqn\Ithreedef{I_3\ =\ {3\pi^4\over\rho^4g^4}}
and we also define $I_2=1$.

To evaluate $I_N$ for general $N$,
it is helpful to rewrite the quadrilinear
term $\Squad$ as a quadratic form. To this end we introduce six independent
auxiliary bosonic variables $\chi^{}_{AB}=-\chi^{}_{BA},$ and substitute into
Eq.~\INdef\ the integral representation
\eqn\auxdef{e^{-\Squad}\ =\ 
{i\rho^6g^6\over2^9\pi^6}\int\prod_{1\le A'<B'\le4}\,d\chi^{}_{A'B'}\,
\exp\big({\rho^2g^2\over32\pi^2}\epsilon^{}_{ABCD}\chi^{}_{AB}\chi^{}_{CD}
+{1\over2}\,\chi^{}_{AB}\Lambda^N_f(\M^A,\M^B)\big)}
where an appropriate analytic continuation of the integration contours
is understood.
Our strategy is to perform only
the integrations over $\mu^A_{N-2}$ and $\mubar^A_{N-2}$ in Eq.~\INdef,
and thereby relate $I_N$ to $I_{N-1}$. Accordingly we break out these
terms from $\Lambda_f^N\,$:
\eqn\breakout{\Lambda_f^N(\M^A,\M^B)\ =\
\Lambda_f^{N-1}(\M^A,\M^B)\ +\ {1\over2\sqrt{2}}\big(\mubar^A_{N-2}\mu^B_{N-2}-
\mubar^B_{N-2}\mu^A_{N-2}\big)}
The $\{\mu^A_{N-2},\mubar^A_{N-2}\}$ integration in Eqs.~\INdef\ and \auxdef\
brings down a factor of
${1\over64}
\det\,\chi^{}_{AB}$. Next we exploit the fact that
the  determinant of an even-dimensional antisymmetric matrix
is a perfect square:
\eqn\factorprop{\det\,\chi^{}_{AB}\ 
=\ \big({1\over8}\,\epsilon^{}_{ABCD}\chi^{}_{AB}\chi^{}_{CD}
\big)^2}
Since the right-hand side of Eq.~\factorprop\ is proportional to the
square of the first term in  Eq.~\auxdef, 
the result of the
 $\{\mu^A_{N-2},\mubar^A_{N-2}\}$ integration can be rewritten
as a parametric second derivative relating $I_N$ to $I_{N-1}\,$:
\eqn\INrec{I_N\ =\ {1\over4}\,
\pi^4\rho^6g^6{\partial^2\over\partial(\rho^2g^2)^2}\,\big(
(\rho^2g^2)^{-3}I_{N-1}\,\big)}
(The insertion of $(\rho^2g^2)^{-3}$ under the parentheses forces the
derivatives to act on the exponent of Eq.~\auxdef, and not on the
prefactor.)
This recursion relation, combined with the initial condition \Ithreedef,
gives finally
\eqn\INanswer{I_N\ =\ {1\over2}\,(2N-2)!\,
\big({\pi^2\over2\rho^2g^2}\big)^{2N-4}\ .}

Combining Eqs.~\measdef, \insertid\ and \INanswer, we therefore find for
the 1-instanton contribution to the 16-fermion correlator in
$\N=4$ SYM theory:
\eqn\YManswer{\eqalign{\langle\Lambda_{\alpha_1}^1(x_1)\cdots
\Lambda_{\alpha_{16}}^4(x_{16})\rangle\ =\ 
C_N&\int {d^4x_0d\rho\over\rho^5}\int\prod_{A=1}^4d^2\xi^A\,d^2\etabar^A\,
\cr\times\ 
&\big(\xi^1_{\alpha_1}-\etabar_\dgamma^1\sigmabar_{m\alpha_1}^\dgamma
\cdot(x_1^m-x_0^m)\big)K_4(x_0,\rho;x_1,0)
\cr\times\cdots\times\ 
&\big(\xi^4_{\alpha_{16}}-\etabar_\dgamma^4\sigmabar_{m\alpha_{16}}^\dgamma
\cdot(x_{16}^m-x_0^m)\big)K_4(x_0,\rho;x_{16},0)}}
where the overall constant $C_N$ is given by
\eqn\CNdef{C_N\ =\ g^{-24}\,{(2N-2)!\over(N-1)!(N-2)!}\,
2^{-2N+57}\,3^{16}\,\pi^{-10}\ .}
Taking the $N\rightarrow\infty$ limit using Stirling's formula gives
\eqn\CNinfty{C_N\ \longrightarrow\  g^{-24}\,\sqrt{N}\,
2^{55}\,3^{16}\,\pi^{-21/2}\ .}
This is in agreement with the IIB prediction, which can be read off
Eqs.~\corresp-\grnfcn:
\eqn\IIBconst{(\alpha')^{-1}\,e^{-\phi/2}\,f_{16}{\Big|}_{\rm 1\hbox{-}inst}
\ \sim\ g^{-24}\,\sqrt{N} \ ,}
up to an overall numerical constant.
We reiterate that the structural agreement between the IIB and
SYM integrals \grnfcn\ and 
\YManswer, and between the powers of $g$ in Eqs.~\CNdef\ and \IIBconst,
was already noted in the $SU(2)$ analysis of BGKR; the new ingredient
here is the  nontrivial
 agreement in the $N$-dependence between
Eqs.~\CNinfty\ and \IIBconst.

$$\scriptscriptstyle{************************}$$
We are indebted to Tim Hollowood and Giancarlo Rossi for clarifying
discussions, and to Tim Hollowood for useful comments on the draft.
ND, VK and MM acknowledge a NATO Collaborative Research Grant,
ND and VK acknowledge the TMR network grant FMRX-CT96-0012
and
SV acknowledges a PPARC Fellowship for support.

\listrefs
\bye